\documentclass[nofootinbib,showpacs,pra,onecolumn,aps,floatfix]{revtex4}
\usepackage{color}
\usepackage{graphicx}
\begin{document}
\title{Quantum-mechanical effects in a linear time-dependent potential}
\author{S. V. Mousavi}
\email{vmoosavi@qom.ac.ir}
\affiliation{Department of Physics, The University of Qom, P. O. Box 37165, Qom, Iran}
\begin{abstract}
The solution of the time-dependent Schr\"{o}dinger equation is discussed for
a particle confined in half-space $x>0$ with a linear potential $V(x)=Kx$ in the following situations:
(a) sudden removal of the wall and switching on the linear potential $V(x)=Kx$ at $x \leq 0$,
(b) sudden removal of the wall and the potential and
(c) sudden removal of the potential. 
A brief discussion of the quantum statistic is presented.
\end{abstract}
\maketitle
%
%
\section{introduction}
The quantum mechanical effects resulting from time-dependent boundary conditions have been studied by some authors \cite{Mo-PR-1952,GeKa-Sov-1976,GaGo-Zphys-1984, CaGrMu-PR-2009, CaRu-PRA-1997, Go-PRA-2002, BaMaHo-PRA-2002, ManMajHo-PLA-2002, MaHo-Pra-2002, Gr-PhysicaB-1988, GaCaVi-PRA-1999, CaMu-EPL-2006}. Diffraction in time was initially introduced by Moshinsky \cite{Mo-PR-1952} by considering a situation involving a beam of particles impinging from the left on a totally absorbing shutter located at the origin which is suddenly turned off at an instant. In such an example, the transient probability current has a close mathematical resemblance with the intensity of light resulting from the Fresnel diffraction by a straight edge. See \cite{CaGrMu-PR-2009} for a recent review. An interesting feature of the solutions for cutoff initial waves, occurring both in the free case \cite{Ho-book-1993} and in the presence of a potential interaction \cite{CaRu-PRA-1997}, is that, if initially there is a zero probability for the particle to be at $x > 0$, as soon as $t = 0^+$, there is instantaneously, a finite, though very small, probability to find the particle at any point $x > 0$. This non-local behavior of the Schrödinger solution is due to its nonrelativistic nature and not a result of the quantum shutter setup \cite{GaCaVi-PRA-1999}. Application of the Klein$-$Gordon equation to the shutter problem \cite{Mo-PR-1952} shows that the probability density is restricted to the accessible region $x < ct$ ($c$ is the speed of light). 

Gerasimov and Kazarnovskii \cite{GeKa-Sov-1976} confined the initial wave in a finite region by introducing a second shutter at the point $x=a$. Godoy \cite{Go-PRA-2002} pointed out the analogy with Fraunhoffer diffraction in the case of small box (compared to the de Broglie length), and Fresnel diffraction, for larger confinements. 

On the other hand, the phenomenon of quantum superarrivals for the reflection and transmission of a Gaussian wave packet in the case of a rectangular potential barrier while it is perturbed by either reducing or increasing its height has also been considered \cite{BaMaHo-PRA-2002, ManMajHo-PLA-2002, MaHo-Pra-2002}. There exists a finite time interval during which the probability of reflection is larger (superarrivals) when the barrier is lowered compared to the unperturbed case \cite{BaMaHo-PRA-2002}. Similarly, during a certain interval of time, the probability of transmission when the barrier is raised exceeds that for free propagation \cite{ManMajHo-PLA-2002}. 
Such studies suggest that the time-varying boundary conditions can give rise to interesting action-at-a-distance effects in quantum mechanics. 

Exactly solvable models give important points about the quantum many-body physics. The Thonks-Girardeau (TG) gas \cite{Gi-JMP-1960} is a such model, which corresponds to infinitely repulsive bosons in one dimension (1D). This model is exactly solvable via a Fermi-Bose mapping, which relates the TG gas to a system of noninteracting spinless fermions in 1D \cite{Gi-JMP-1960}.

In this context, by considering the problem of a confined particle in a linear potential, the aim of the present paper is to probe some aspects of the time-dependent boundary conditions that have remained hitherto unnoticed. 

The organization of the paper is as follows: In Section \ref{sec: review} we review the solution of the Schr\"{o}dinger equation for a confined particle in half space $0<x$ with the potential $V(x)=Kx$ and consider the various time-dependent boundary condition of the problem. Section \ref{sec: qu_st} briefly deals with quantum statistics. Numerical calculations are presented in section \ref{sec: nuca}. Section \ref{sec: sudi} contains concluding remarks. 
\section{A confined particle in half space $0<x$ with the potential $V(x)=Kx$} \label{sec: review}
\subsection{Stationay states}
The Schr\"{o}dinger equation of the $n^{th}$ state of the particle wave function is
\begin{eqnarray}\label{eq: Sch}
\frac{d^2 \phi_n(x)}{dx^2}+\frac{2m}{\hbar^2}(E_n-Kx)\phi_n(x) &=& 0~.
\end{eqnarray}
Performing the change of variable
\begin{eqnarray}
\xi &=& (\frac{E_n}{K}-x) \alpha \nonumber\\
\alpha &=& \left( \frac{2mK}{\hbar^2} \right)^{1/3}~,
\end{eqnarray}
we get the Airy equation,
\begin{eqnarray}
\frac{d^2 \phi_n}{d\xi^2} + \xi \phi_n &=& 0~,
\end{eqnarray}
whit solutions,
\begin{eqnarray}
\phi_n(x) &=& N_n Ai(-\xi)~.
\end{eqnarray}
in which $N_n$ is the normalization constant.
The energy levels are determined by the condition $\phi_n(0)=0$
\begin{eqnarray}
Ai\left[- \alpha \frac{E_n}{K} \right] &=& 0 ~.
\end{eqnarray}
Then, the $E_n$ are determined by the zeros $a_n$ of the Airy function
\begin{eqnarray}
E_n &=& -a_n \frac{K}{\alpha} = -a_n \left( \frac{\hbar^2 K^2}{2m} \right)^{1/3} ~.
\end{eqnarray}
Thus, the solution of Eq. (\ref{eq: Sch}) is given by \cite{VaSo-book-2004},
\begin{eqnarray} \label{eq: initial_wave}
\phi_n(x )&=& \sqrt{\alpha} \frac{1}{Ai^{\prime}(a_n)} Ai (\alpha x+a_n)~,
\end{eqnarray}
in which ${Ai^{\prime}(a_n)} = \frac{dAi(x)}{dx}|_{x=a_n}$. 
Now, we study the evolution of Eq. (\ref{eq: initial_wave}) under time-dependent Schr\"{o}dinger equation in the following cases:
%
%
\subsection{Sudden removal of the wall and switching on the linear potential $V(x)=Kx$ at $x \leq 0$}
Suddenly the wall at $x=0$ is removed at time equal to zero, and a linear potential of the form $V(x)=Kx$ is switched on in the region $x \leq 0$.
Now, the particle is in a linear potential in whole space.
At any instant $t>0$ wavefunction is obtained by,
\begin{eqnarray} \label{eq: wf(t)}
\psi(x, t) &=& \int_{-\infty}^{\infty} dx^{\prime}~ G(x, t|x^{\prime}, 0)\psi_0(x^{\prime})~,
\end{eqnarray}
in which $\psi_0(x)$ is the initial wavefunction and is chosen as Eq. (\ref{eq: initial_wave}) and $G(x, t|x^{\prime}, 0)$ is the propagator for the Hamiltonian $H = p^2/2m+Kx$ and is given by \cite{GrSt-book-1998},
\begin{eqnarray} \label{eq: prop-li}
G(x, t|x^{\prime}, 0) &=& \sqrt{\frac{m}{2\pi i \hbar t}} ~\text{e}^{\frac{im}{2\hbar t} (x-x^{\prime})^2-\frac{iKt}{2\hbar}(x+x^{\prime})-\frac{iK^2 t^3}{24\hbar}}~.
\end{eqnarray}
Putting Eq. (\ref{eq: prop-li}) into Eq. (\ref{eq: wf(t)}), one finds
\begin{eqnarray}
\psi_n(x, t) &=& \sqrt{\alpha} \frac{1}{Ai^{\prime}(a_n)} ~\text{e}^{-\frac{iKt}{2\hbar}x-\frac{iK^2 t^3}{24\hbar}} 
\sqrt{\frac{m}{2\pi i \hbar t}} \int_0^{\infty} dx^{\prime} ~\text{e}^{\frac{im}{2\hbar t} (x-x^{\prime})^2 -\frac{iKt}{2\hbar}x^{\prime}} Ai(\alpha x^{\prime}+a_n)~.
\end{eqnarray}
Inverse Fourier transform of the function $\tilde{Ai}(k)=~\text{e}^{ik^3/3}$ is the Airy function,
\begin{eqnarray}
Ai(x) &=& \frac{1}{2\pi}\int_{-\infty}^{\infty} dk ~\text{e}^{ikx}\tilde{Ai}(k)~.
\end{eqnarray}
So,
\begin{eqnarray} \label{eq: wf_li}
\psi_n (x, t) &\equiv& \psi_n^{\text{(a)}}(x, t) = \frac{\sqrt{\alpha}}{2\pi} \frac{1}{Ai^{\prime}(a_n)} ~\text{e}^{-\frac{iKt}{2\hbar}x-\frac{iK^2 t^3}{24\hbar}} 
\sqrt{\frac{m}{2\pi i \hbar t}}
\int_{-\infty}^{\infty} dk ~\text{e}^{ik^3/3} ~\text{e}^{ika_n} \int_0^{\infty} dx^{\prime} ~\text{e}^{\frac{im}{2\hbar t} (x-x^{\prime})^2-\frac{iKt}{2\hbar}x^{\prime}} ~\text{e}^{i \alpha k x^{\prime}} \nonumber\\
&=&
\frac{1}{2\pi\sqrt{\alpha}} \frac{1}{Ai^{\prime}(a_n)} ~\text{e}^{-\frac{iKt}{2\hbar}x-\frac{iK^2 t^3}{24\hbar}} 
\int_{-\infty}^{\infty} dk ~\text{e}^{ik^3/3\alpha^3} ~\text{e}^{ia_nk/\alpha} \sqrt{\frac{m}{2\pi i \hbar t}} \int_0^{\infty} dx^{\prime} ~\text{e}^{\frac{im}{2\hbar t} (x-x^{\prime})^2-\frac{iKt}{2\hbar}x^{\prime}} ~\text{e}^{i k x^{\prime}} \nonumber\\
&=&
\frac{1}{4\pi\sqrt{\alpha}} \frac{1}{Ai^{\prime}(a_n)} ~\text{e}^{-\frac{iKt}{2\hbar}x-\frac{iK^2 t^3}{24\hbar}} 
\int_{-\infty}^{\infty} dk ~\text{e}^{ik^3/3\alpha^3} ~\text{e}^{ia_nk/\alpha} ~\text{e}^{\frac{im}{2\hbar t}(x^2-X^2)} \text{erfc}\left[ \frac{i+1}{2} \sqrt{\frac{m}{\hbar t}} X \right]~,
\end{eqnarray}
in which $X=x+Kt^2/2m-\hbar kt/m$ and erfc$(x)$ is the complementary error function.
%
%
\subsection{Sudden removal of the wall and the potential (Free propagation)}
By sudden removal of the wall and turning off the potential at time equal to zero, the particle is allowed to propagate freely.
Using Eq. (\ref{eq: prop-li}) for $K=0$, one finds
\begin{eqnarray} \label{eq: wf_fr}
\psi_n^{\text{(b)}}(x, t) &=& 
\frac{1}{4\pi \sqrt{\alpha}} \frac{1}{Ai^{\prime}(a_n)}
\int_{-\infty}^{\infty} dk ~\text{e}^{i k^3/3\alpha^3 -iEt/\hbar + ik(a_n/\alpha + x)}
\text{erfc} \left[\frac{i+1}{2} \sqrt{\frac{m}{\hbar t}} 
(x-vt)\right]~,
\end{eqnarray}
in which, $E = \hbar^2 k^2/2m$ and $v =\hbar k/m$. From Eqs. (\ref{eq: wf_li}) and (\ref{eq: wf_fr}) one gets the following relation between the density probability in the free and the interacting case:
\begin{eqnarray} \label{eq: free-linear}
|\psi_n^{\text{(a)}}(x, t)|^2 &=& |\psi_n^{\text{(b)}}(x+\frac{Kt^2}{2m}, t)|^2~.
\end{eqnarray}
Thanks to this relation we do numerical calculations only for the free case. 
%
%
\subsection{Sudden removal of the potential}
In this case the propagator is given by \cite{Kl-PR-1994},
\begin{eqnarray} 
G(x, t|x^{\prime}, 0) &=& G_{\text{free}}(x, t|x^{\prime}, 0)-G_{\text{free}}(-x, t|x^{\prime}, 0)~,
\end{eqnarray}
where $G_{\text{free}}(x, t|x^{\prime}, 0)$ denotes the propagator of free particle and is given by Eq. (\ref{eq: prop-li}) with $K=0$.
From Eq. (\ref{eq: wf_fr}), we have 
\begin{eqnarray} 
\psi_n^{\text{(c)}}(x, t) &=& 
\frac{1}{4\pi \sqrt{\alpha}} \frac{1}{Ai^{\prime}(a_n)}
\int_{-\infty}^{\infty} dk ~\text{e}^{i k^3/3\alpha^3 -iEt/\hbar + ika_n/\alpha} \nonumber\\
&\times&
\left( ~\text{e}^{ikx} \text{erfc} \left[\frac{i+1}{2} \sqrt{\frac{m}{\hbar t}} (x-vt)\right] - ~\text{e}^{-ikx} \text{erfc} \left[\frac{i+1}{2} \sqrt{\frac{m}{\hbar t}} (-x-vt)\right] \right)~.
\end{eqnarray}
%
\section{Quantum Statistics} \label{sec: qu_st}
So far we have not consider quantum statistics. We shall next consider dynamics of the ground state of a system of $N$ noninteracting spinless Fermions initially confined in half-space $x>0$ in the presence of a linear potential. Since wavefunctions of noninteracting spinless fermions are antisymmetric under coordinate exchanges, their wavefunctions vanish automatically whenever any $x_j=x_k$:
\begin{eqnarray}
\psi_{F0}(x_1, ..., x_N, t) &=& \frac{1}{\sqrt{N!}} \text{det}_{(n,j)=(1,1)}^{(N,N)} \psi_n(x_j, t)~.
\end{eqnarray}
The single particle density, normalized to $N$, is
\begin{eqnarray}
\rho(x, t) &=& N \int |\psi_{F0}(x, x_2, ..., x_N, t)|^2~dx_2...dx_N = \sum_{n=1}^{N} |\psi_n (x, t)|^2.
\end{eqnarray}
It follows from the Fermi-Bose mapping theorem \cite{Gi-JMP-1960, Gi-PRA-1965, GiWr-PRL-2000} that the exact $N-$bosons ground state wavefunction $\psi_{B0}$ of a system of $N$ bosons with hard-wall interaction is
\begin{eqnarray}
\psi_{B0}(x_1, ..., x_N, t) &=& |\psi_{F0}(x_1, ..., x_N, t)|~.
\end{eqnarray}
%
\section{Numerical calculations} \label{sec: nuca}
In this section we work in a unit system where $\hbar=1$ and $m=0.5$. We choose $K=1$ anywhere the potential is nonzero. Thus, from Eq. (\ref{eq: initial_wave}) the initial wavefunction is given by, 
\begin{eqnarray}
\psi_0(x) &\equiv& \phi_n(x ) = \frac{1}{Ai^{\prime}(a_n)} Ai(x+a_n)~.
\end{eqnarray}
We use Kr\"{u}ger's method \cite{Kr-Theo-1981, VaSo-book-2004} to produce Airy functions.
\subsection{Sudden removal of the wall and the potential (Free propagation)}
\begin{figure}
\centering
\includegraphics[width=12cm,angle=0]{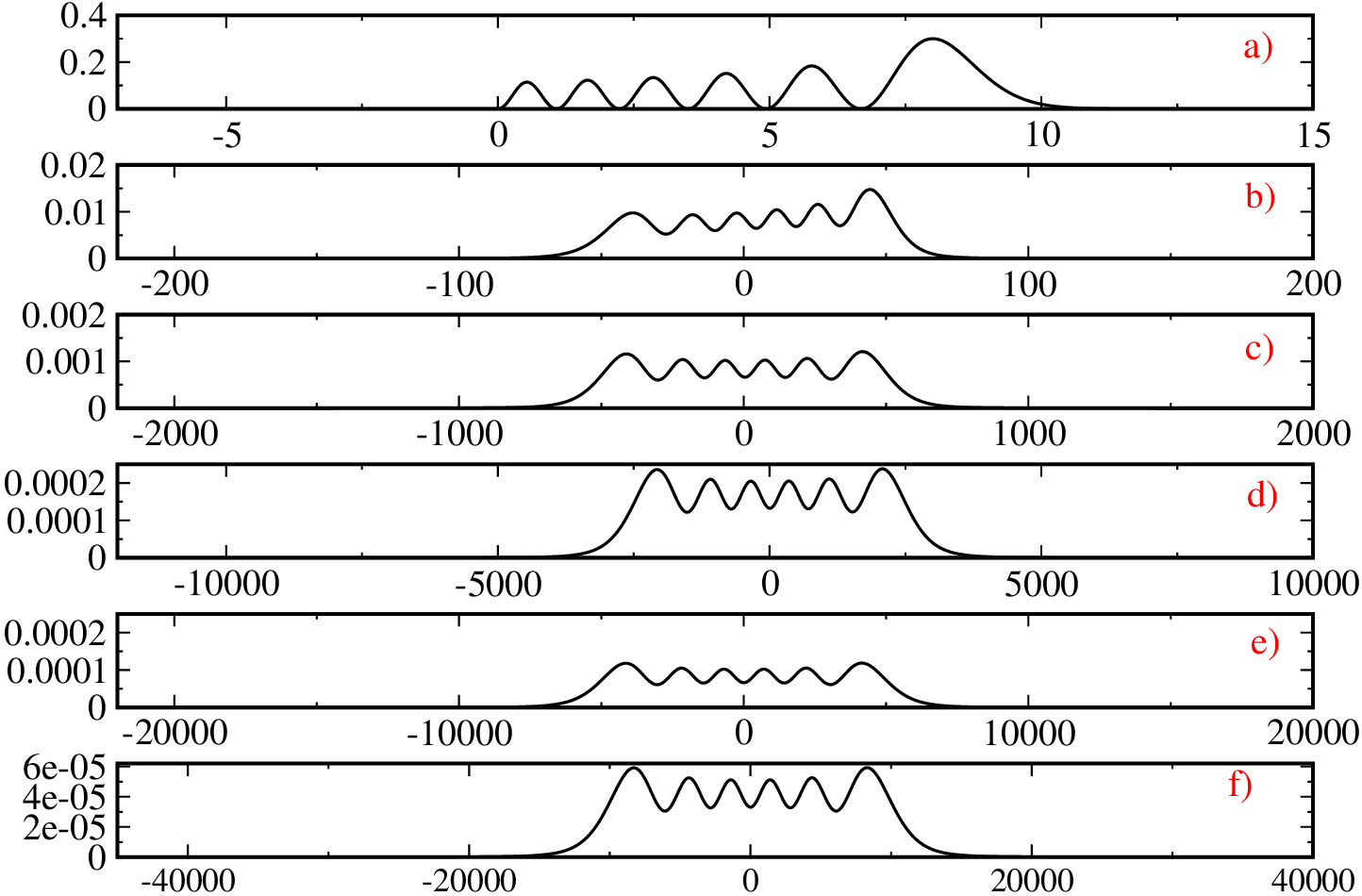}
\caption{Probability density versus distance $x$ for state $n=6$ at times (a) $t=0$, (b) $t=10$, (c) $t=100$, (d) $t=500$, 
(e) $t=1000$ and (f) $t=2000$. (Free propagation)}
\vspace*{1cm}
\label{fig: denx_b}
\end{figure}

\begin{figure}
\centering
\includegraphics[width=10cm,angle=-90]{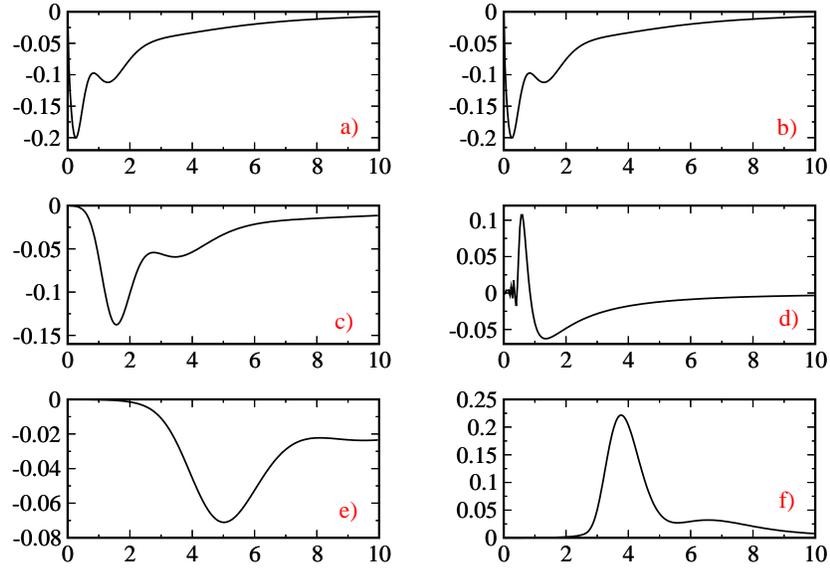}%
\caption{Probability current density versus time $t$ for state $n=6$ at points (a) $x=-0.0001$, (b) $x=0.0001$, (c) $x=-5$, (d) $x=5$, (e) $x=-20$ and (f) $x=20$. (Free propagation)}
\label{fig: curt_b}
\end{figure}

\begin{figure}
\centering
\includegraphics[width=10cm,angle=-90]{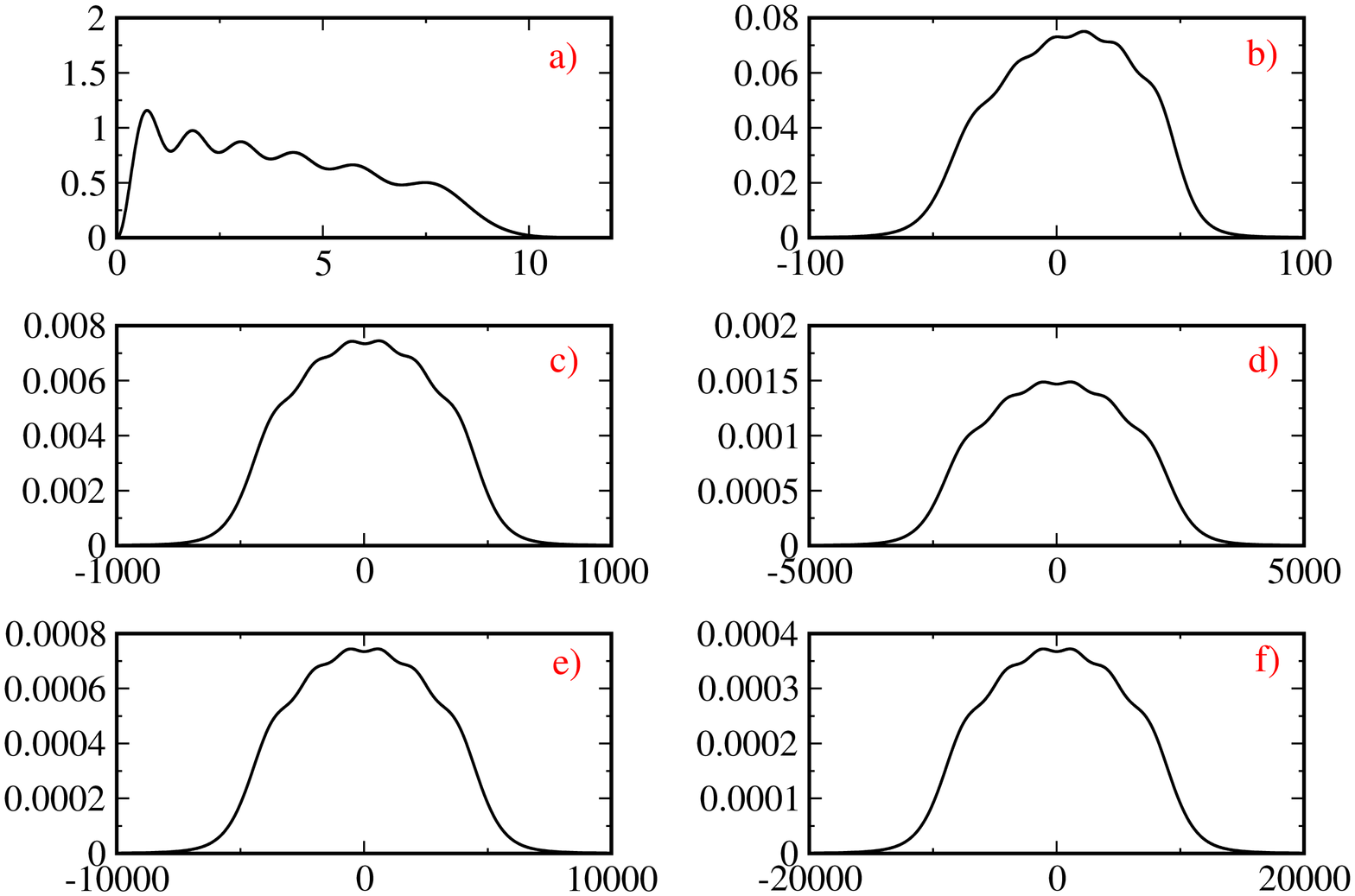}%
\caption{Density plots of the probability density $\rho=\sum_{n=1}^N |\psi_n(x, t)|^2$ for a noninteracting spinless Fermioninc gas composed of $N=6$ particles versus distance $x$ at times (a) $t=0$, (b) $t=10$, (c) $t=100$, (d) $t=500$, (e) $t=1000$ and (f) $t=2000$. (Free propagation)}
\label{fig: denprofile_b}
\end{figure}

Fig. \ref{fig: denx_b} shows the probability density versus distance $x$ for state $n=6$ with energy eigenvalue $E_6=9.023$ at different times. From this figure one can see that at long times probability density locates symmetrically around the origin and it spreads without motion. 
From Eqs. (\ref{eq: wf(t)}) and (\ref{eq: prop-li}) for free propagation one has,
\begin{eqnarray} 
\psi(x, t) &=& \sqrt{\frac{m}{2\pi i \hbar t}} \int_0^{\infty} dx^{\prime}~\text{e}^{\frac{im}{2\hbar t} (x-x^{\prime})^2} \phi_n(x^{\prime}) = \sqrt{\frac{m}{2\pi i \hbar t}}~\text{e}^{\frac{im}{2\hbar t} x^2}
\int_0^{\infty} dx^{\prime}~\text{e}^{-\frac{im}{\hbar t}xx^{\prime}} \text{e}^{\frac{im}{2\hbar t}x^{\prime 2}}\phi_n(x^{\prime})~,
\end{eqnarray}
and
\begin{eqnarray}
\rho(x, t) &=& \frac{m}{2\pi \hbar t} \left\vert\int_0^{\infty} dx^{\prime}~ \text{e}^{-\frac{im}{\hbar t}xx^{\prime}} \text{e}^{\frac{im}{2\hbar t} x^{\prime 2}}\phi_n(x^{\prime})\right\vert^2 \nonumber\\
&=& 
\frac{m}{2\pi \hbar t} \left\vert\int_0^{\infty} dx^{\prime}~ \cos\left( {\frac{mx}{\hbar t}x^{\prime}}\right) \text{e}^{\frac{im}{2\hbar t} x^{\prime 2}}\phi_n(x^{\prime})\right\vert^2 + \frac{m}{2\pi \hbar t} \left\vert\int_0^{\infty} dx^{\prime}~ \sin\left( {\frac{mx}{\hbar t}x^{\prime}}\right) \text{e}^{\frac{im}{2\hbar t} x^{\prime 2}}\phi_n(x^{\prime})\right\vert^2 
\nonumber\\
&+& \frac{m}{2\pi \hbar t} \left[- i
\int_0^{\infty} dx^{\prime}~ \cos\left( {\frac{mx}{\hbar t}x^{\prime}}\right) \text{e}^{-\frac{im}{2\hbar t} x^{\prime 2}}\phi_n(x^{\prime}) \times \int_0^{\infty} dx^{\prime \prime}~ \sin\left( {\frac{mx}{\hbar t}x^{\prime \prime}}\right) \text{e}^{\frac{im}{2\hbar t} x^{\prime \prime 2}}\phi_n(x^{\prime \prime}) +\text{c.c.}
\right]~,
\end{eqnarray}
in which c.c. refers to the complex conjugate of the first term in the bracket.
One can check numerically in the limit $t\rightarrow \infty$, last line of the above expression is very small compared to the two first terms. As a result of which density probability would be an even function of $x$ at long times. 
After a certain time $t_n$, for even $n$, $|\psi_n(x, t)|^2$ has a local minimum at $x=0$ and its $n$ maxima 
are distributed symmetrically around $x=0$, but for odd $n$ it has a local maximum at $x=0$ and other $n-1$ maxima locate
symmetrically around the central maximum, i.e., $(n-1)/2$ maxima at the right and $(n-1)/2$ maxima at the left of the origin. Another feature of 
$|\psi_n(x, t)|^2$ is that it has no node after a certain time of letting the particle to be free.
From Eq. (\ref{eq: free-linear}) one can find this is also true for propagating of $\phi_n(x)$ in a linear potential. 
Another physically interesting quantity is the probability current density $j=(\hbar/m) \Im (\psi^* \partial \psi/\partial x)$. It is important because arrival time distribution at a given location is given by its modulus sign in some approaches to this problem \cite{MuLe-PR-2000}.
Fig. \ref{fig: curt_b} shows the probability current density versus time at different locations. From this figure one can see $j$ is negative for negative values of $x$ and is positive for large positive values of $x$, but its sign changes with time for small values of positive $x$. 
Fig. \ref{fig: denprofile_b} represents the evolution of the density profile for a noninteracting spinless Fermioninc gas composed of $N=6$ atoms released from the trap. When the particle is trapped in the $n-$th mode, and after being released, the probability density presents $n$ maxima and by elapsing the time, they distribute symmetrically around the origin. As a result of which, for the case of $N-$body system, single particle density presents $N$ maxima, when particles are trapped and after being released and by passing the time these maxima locates symmetrically around the origin.
\subsection{Sudden removal of the potential}
\begin{figure}
\centering
\includegraphics[width=10cm,angle=0]{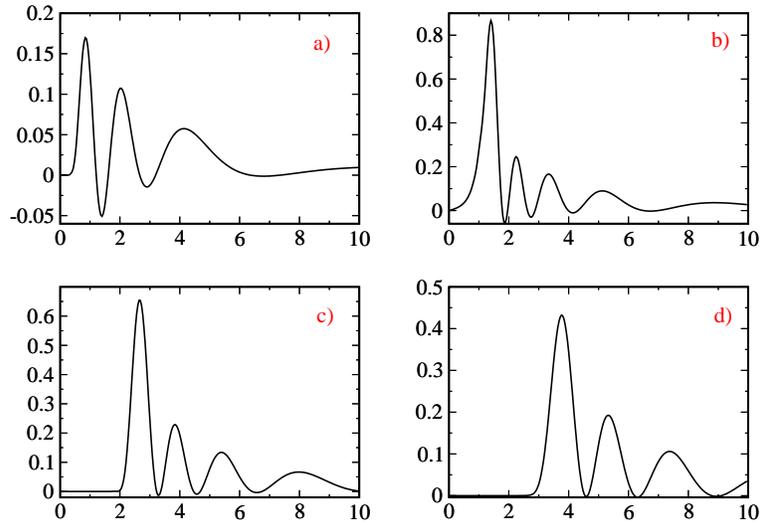}%
\caption{Probability current density versus time $t$ for state $n=6$ at points (a) $x=5$, (b) $x=10$, (c) $x=15$ and (d) $x=20$. (Free propagation in half-space $x>0$)}
\label{fig: curt_c}
\end{figure}

\begin{figure}
\centering
\includegraphics[width=10cm,angle=-90]{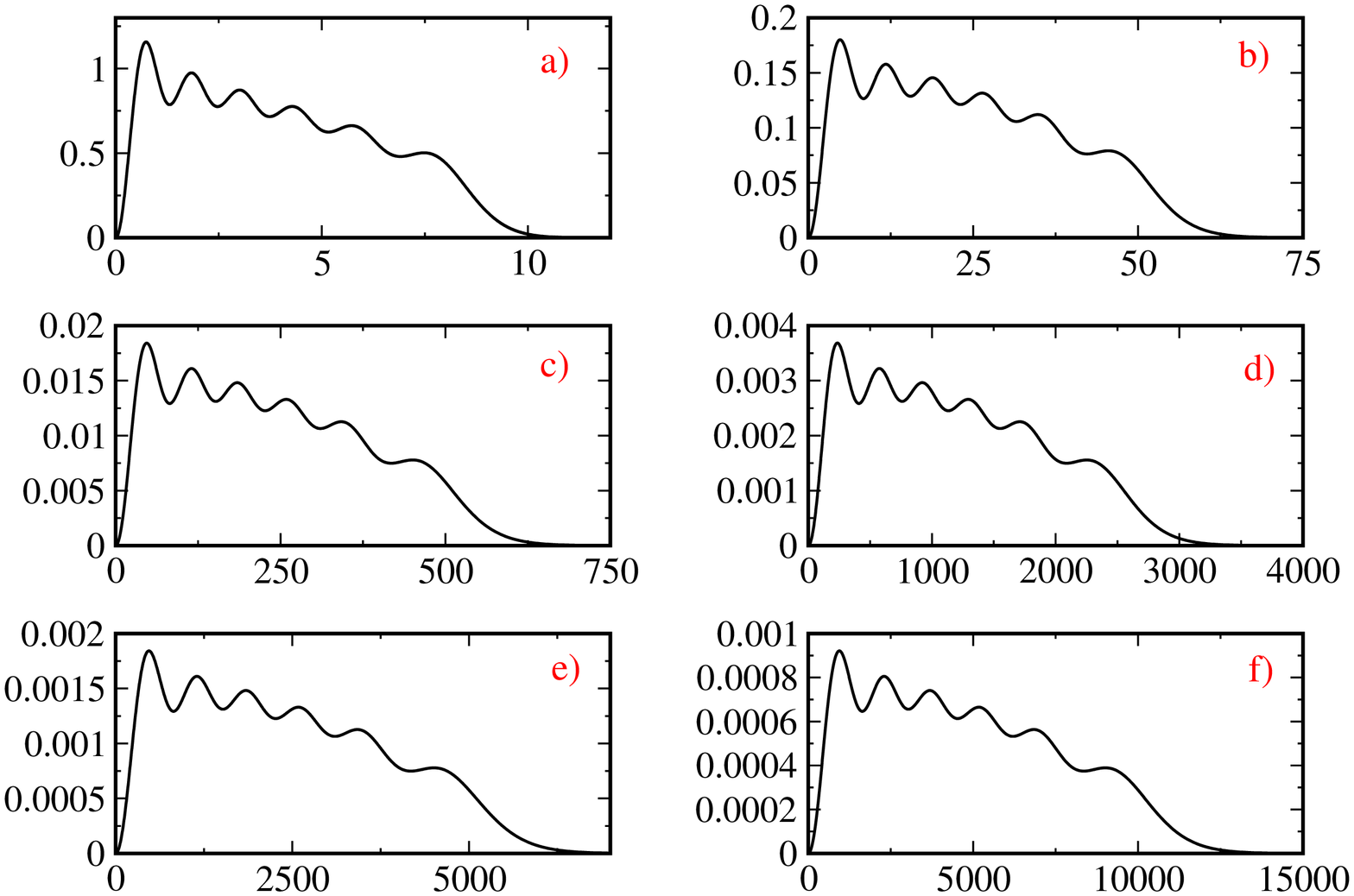}%
\caption{Density plots of the probability density $\rho=\sum_{n=1}^N |\psi_n(x, t)|^2$ for a noninteracting spinless Fermioninc gas composed of $N=6$ particles versus distance $x$ at times (a) $t=0$, (b) $t=10$, (c) $t=100$, (d) $t=500$, (e) $t=1000$ and (f) $t=2000$. (Free propagation in half-space $x>0$)}
\label{fig: denprofile_c}
\end{figure}

Fig. \ref{fig: denx_c} shows probability density versus distance $x$ for state $n=6$ at different times. It cen be seen that wavefunction spreads without motion. The height of the last maximum is larger than the $n-1$ first ones before turning the potential off, but by going time, the height of them goes to be the same except the last one, whose height is smaller than the others.
$|\psi_n(x, t)|^2$ in Fig. \ref{fig: denx_c} has always $n$ nodes in spite of Fig. \ref{fig: denx_b}. This comes from the fact that in this case the wall is in place.
Fig. \ref{fig: curt_c} shows the probability current density versus time at different locations. From this figure one can see that $j$ is positive for large values of $x$, but its sign changes with time for small values of $x$. 
Finally, Fig. \ref{fig: denprofile_c} represents the evolution of the density profile for a Fermioninc gas composed of $N=6$ noninteracting spinless atoms released from the trap. Noting this figure, one sees the height of the maxima decreases by $x$ for a given time.
\section{Sumarry and discussion} \label{sec: sudi}
In this paper we have considered numerical solution of the time-dependent Schr\"{o}dinger equation for some time-dependent boundary conditions. Probability density, probability current density and dynamics of the ground state of a system of $N$ noninteracting spinless Fermions were studied. When the particle is trapped in the $n-$th mode, and after being released, the probability density presents $n$ maxima. As a result of which, for the case of $N-$body system, single particle density presents $N$ maxima, when particles are trapped and after being released.
Since the stationary wavefunctions $\phi_n(x)$ given in Eq. (\ref{eq: initial_wave}) form a complete set of orthonormal eigenfunctions in the range $(0 \leq x)$, we can expand an initial wave packet $\psi(x, 0)$ which has an arbitrary shape for $(0 \leq x)$ in terms of $\phi_n(x)$
\begin{eqnarray} \label{eq: expansion}
\psi(x, 0) &=& \sum_{n=0}^{\infty} c_n \phi_n(x)~,
\end{eqnarray}
where the expansion coefficients $c_n$ is given by: $c_n=\int_0^{\infty} \phi_n^*(x) \psi(x, 0)~dx$. We have find time evolution of eigenfunctions $\phi_n(x)$ which were denoted by $\psi_n(x, t)$. Thus, time evolution of Eq. (\ref{eq: expansion}) is given by 
\begin{eqnarray}
\psi(x, t) &=& \sum_{n=0}^{\infty} c_n \psi_n(x, t)~.
\end{eqnarray}
Time-dependent probability density $|\psi(x, t)|^2$ is a complicated linear combination of $|\psi_n(x, t)|^2$ and the corresponding interferences amplitudes $\psi_n^*(x, t) \psi_{n^{\prime}}(x, t)$.
%

\pagebreak

\begin{figure}
\centering
\includegraphics[width=12cm,angle=0]{wall_denx.eps}%
\caption{Probability density versus distance $x$ for state $n=6$ at times (a) $t=0$, (b) $t=10$, (c) $t=100$, (d) $t=500$, 
(e) $t=1000$ and (f) $t=2000$. (Fee propagation in half-space $x>0$)}
\label{fig: denx_c}
\end{figure}
\end{document}